# Direct observation of the reversible and irreversible processes in femtosecond laser-irradiated fused silica at the near-damage-threshold fluence

## Shaohua Ye and Min Huang*

*State Key Laboratory of Optoelectronic Materials and Technologies and School of Physics, Sun Yat-Sen University, Guangzhou, 510275, China*
*Corresponding author: syshm@163.com*

For fused silica irradiated by near-100-fs, 795-nm laser pulses with fluence approaching the damage threshold, the transient transmission spectroscopy based on a wavelength-degenerate pump-probe technique clearly presents two dynamic processes corresponding to the instantaneous effects of laser optical field and the delayed effects of free electron dynamics, respectively. The reversible, instantaneous process originates in third-order nonlinear optical responses (in particular the optical Kerr effect) ascribed to virtual optical-field ionization (VOFI) that significantly contributes to the nonlinear optical polarization with energy exchange recoverability. Whereas, the irreversible, delayed process originates in the effects of free electron plasma generated by initial real optical-field ionization (ROFI) and subsequent impact ionization (II), being responsible for the energy dissipation and optical breakdown. In general, the femtosecond wavelength-degenerate pump-probe spectroscopy can detect VOFI, ROFI, and II simultaneously in strong-field nonlinear polarization and ionization of fused silica, and offer flexible ways to distinguish the different mechanisms. For the near-100-fs pulses, our results confirm that II provided with a typical delay time about 300 fs is responsible for the optical breakdown of fused silica.

## 1. Introduction

In general, femtosecond laser irradiation of dielectrics at the near-damage-threshold fluence will initially generate massive free electrons via optical-field ionization (OFI) (two extreme limits are multiphoton ionization and tunneling ionization in term of the Keldysh theory of strong-field photoionization). Then, in the strong optical field the excited free electrons can further gain kinetic energy through inverse bremsstrahlung absorption (IBA) and be heating continuously. After adequate heating certain free electrons will obtain the threshold kinetic energy to excite other bound electrons via impact ionization (II)—a mechanism may eventually cause severe avalanche breakdown in the dielectric. Theoretically, for the highly-nonlinear characteristics of OFI and II, the roles of these two ionization mechanisms depend on many factors, such as pulse duration, irradiation fluence, material property, surface morphology, etc. Therefore, the strengths of the two ionization processes cannot easily be determined for a material under certain irradiation conditions [1-15]. In fact, due to the difference in the theoretical frameworks, material parameters, and irradiation conditions, distinct conclusions would be obtained. For instance, the studies on ultrafast laser ionization of dielectrics show that II may play an important role [16-24] or an inappreciable role [25-28] in the similar irradiated condition. Hence, more affirmative conclusion on the issue depends on the obtaining of more direct and explicit experimental evidence for the transient ionization processes.

Up to now, it is still very challenging for researchers to distinguish between OFI and II in solids through a certain transient measurement technique directly. Generally speaking, by the traditional strong-field pump-probe detection, one is easy to observe some effects caused by ultrafast laser ionization of solids, such as the decrease or increase of the transient transmissivity or reflectivity. Nevertheless, the typical features of the transient spectra are often in the time scale of a few hundred femtoseconds (fs), which is significantly greater than the pulse duration and only reflects the signal mixing the effects of OFI and II [23,24,28-49]—it seems that the time resolution and signal-to-noise ratio of the traditional pump-probe technique is not high enough to clearly distinguish between OFI and II. But it is noteworthy that some reported transient spectra showed certain blurring features in time scale close to pulse duration [32,47-49]. These results imply that transient measurements provided with higher time resolution and signal accuracy would bring us more direct and detailed evidences about the dynamics and roles of OFI and II, and thus provide the possibility of verifying the stage characteristics of time evolution of theoretical transient spectra for different evolutionary processes between OFI and II [12,15]. Recently, the third-order nonlinear optical effects (TNOEs), in particular the optical Kerr effect (OKE) that is always accompanied with ionization effects and provided with an instantaneous response characteristic, has been paid attention to [23,24,48] and used as an auxiliary signal in transient measurements for accurately determining the pump-probe time zero [24,48]. These studies would provide us a new idea to improve the traditional strong-field pump-probe technique.

On the other hand, in recent years, with the rapid progress of attosecond detection technology, researchers are able to probe into the wave-cycle-resolved strong-field-induced effects in dielectrics [50-56]. These inspiring results reveal a reversible characteristic of the field-induced changes of typical physical quantities with laser intensity close to the optical breakdown threshold. Significantly, strong-field excitation of electrons appears a strong virtual characteristic [53,55-59]: after the impinging of a laser pulse, a large part of the conduction-band population created directly by the instantaneous field will return to valence band, acting as a kind of virtual population provided with the energy reversibility, and the residual part of the conduction-band population will keep on living as a kind of real population causing the energy dissipation. In the traditional theory of strong-field nonlinear polarization ($P_{NL}$) of dielectrics, $P_{NL} = P_{bound} + P_{free}$, comprising the responses of bound and free electrons (the response of bound electrons are further modelled via instantaneous Kerr and delayed Raman responses). Accordingly, the virtual population contributes to $P_{bound}$ instantaneously [54] and thus should enhance OKE of TNOEs prominently, whereas the real population contributes to $P_{free}$ that may suppress and saturate OKE [60,61]. With regard of the virtual population, conceptually the traditional OFI can be extended as the virtual OFI (VOFI) associated to the virtual population and the real OFI (ROFI) associated to the real population: VOFI contributes to the nonlinear polarization, in particular OKE of TNOEs, and ROFI contribute to the saturation of OKE and the further induction of II; with laser fluence approaching the damage threshold, the ratio of ROFI to VOFI increases rapidly—a critical physical regime turning from VOFI to ROFI. In short, the attosecond streaking spectrogram with the wave-cycle-resolved ability, a very powerful tool for studying the instantaneous responding of materials to the strong fields, would provide new insights for us to understand the strong-field nonlinear polarization and ionization of solids. Noteworthy, II and the relationship between ROFI and II have not been touched upon by the attosecond detection technology, which is worthy of in-depth exploration.

As introduced above, the dominant ionization mechanism causing the optical damage for a certain solid irradiated by femtosecond laser pulses is still an unresolved fundamental issue in the field of ultrafast laser-solid interaction [16-28]. In the paper, we develop a wavelength-degenerate pump-probe technique provided with high beam-divergence-state sensitivity as well as high signal-to-noise ratio and time resolution. For fused silica irradiated by near-100-fs, 795-nm laser pulses with fluence approaching the damage threshold, the femtosecond wavelength-degenerate pump-probe transmission spectroscopy (FWPTS) can clearly present two dynamic processes, which are related to the instantaneous effects of laser optical field and the delayed effects of free electron dynamics, respectively. In fact, the transient spectrum studies based on the wavelength-degenerate pump-probe technique can provide a straightforward way for the comprehensive detection of VOFI, ROFI, and II simultaneously in strong-field nonlinear polarization and ionization of fused silica, and give insights into the dominant ionization mechanism for ultrafast laser-induced damage of fused silica, which has not been completely covered by the previous femtosecond and attosecond strong-field transient spectroscopies.

## 2. Experimental details

### 2.1 The basic experimental setup of the wavelength-degenerate pump-probe measurement

In the study, an optical polishing fused silica sheet (KMT, 1-inch diameter, 1-mm thickness, and < 5 Å surface roughness) is used as the sample. The schematic diagram of experimental setup for the wavelength-degenerate pump-probe measurement is shown in Fig. 1(a). A Ti:sapphire laser system (COHERENT, Legend Elite USP HE+) with a 795-nm central wavelength, a 1-kHz repetition rate, and a horizontal linear polarization is utilized as the intense pulsed light source, which can provide ultrashort pulses with a full-width-at-half-maximum (FWHM) pulse duration near 100 fs at the pump-probe system measured by the autocorrelation technique.

Then, the main beam is split into a strong pump beam and a weak probe beam with the respective pulse energies precisely controlled by circular variable neutral density filters, and the respective polarizations controlled by half wave plates. A computer-controlled stepper-motor-driven delay line is used to adjust the time delay between the pump and probe pulses with a displacement resolution of 1.25 μm, corresponding to a time resolution accuracy of 8.3 fs. The pump beam along the surface normal and the probe beam along the 30-degree-angle direction are focused on the sample front surface of the same location by convex lenses with the focal lengths of 10 cm and 5 cm and the focal spot diameters of about 20 μm and 10 μm (FWHM), respectively; the 30-degree intersection for the two beams is large enough to eliminate the influence of the pump stray light on the probe beam receiving system. The surface appearance of the irradiated region and the overlapped situation of the two focal spots are monitored by a CCD camera with the help of a high magnification zoom lens. The transparent fused silica sheet is mounted on a 3D micro displacement platform for ease of adjustment of the irradiation spot and the focusing situation: the measurement of each fluence series is launched at a fresh irradiation spot.

Finally, the transmitted probe beam is detected by a signal acquisition system with a light-collecting module sensitive to the divergence state of probe beam, as described detailedly in the following C part. In the light intensity detection mode of the signal acquisition system, a Si photodiode detector (Thorlabs, S120C) controlled by an optical power meter (Thorlabs, PM320E) is used for detecting the intensity of the transmitted probe beam. For improving the signal-to-noise ratio of the detection, the signal acquisition process is based on the lock-in-amplification technique with the help of an optical chopper (Stanford Research Systems, SR 540) used in the light path of the probe beam and the lock-in amplifier (Stanford Research Systems, SR 830) used in the signal detection terminal. The whole signal acquisition system is controlled by a self-developed software to realize automatic FWPTS measurement with high precision and repeatability. In the spectral measurement mode of the signal acquisition system, the wavelength spectrum of the transmission probe beam for a specific delay time is measured by a fiber optic spectrometer (Ocean Optics, HR4000) equipped with an integrating sphere (Ocean Optics, ISP-REF) for collecting the beam.

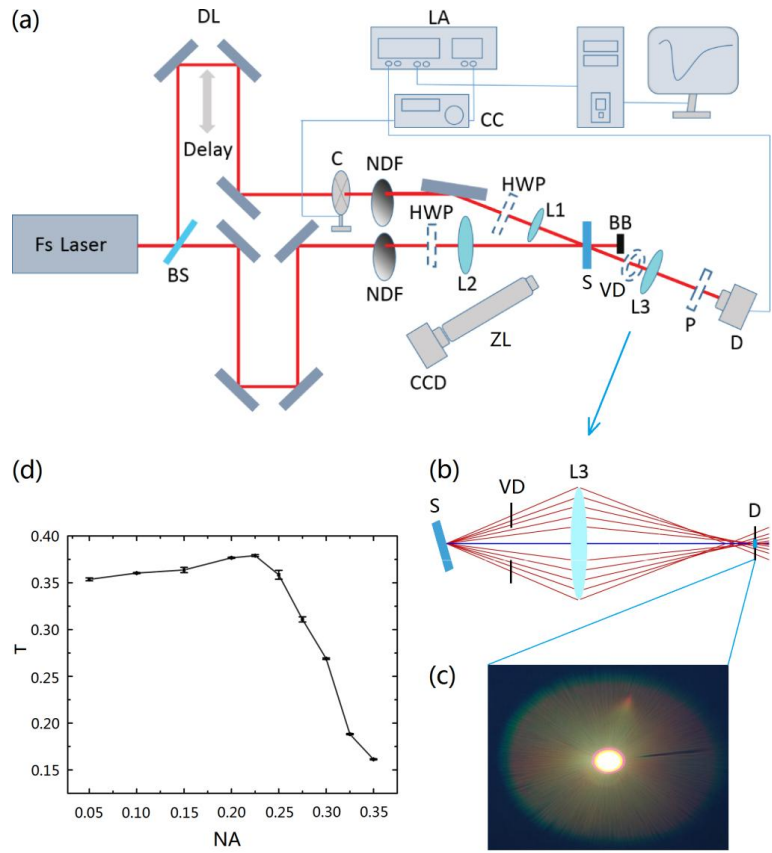

Fig. 1. (a) The schematic diagram of experimental setup for FWPTS measurement, which includes the femtosecond laser (Fs laser), beam splitter (BS), delay line (DL), optical chopper (OC), chopper controller (CC), lock-in amplifier (LA), circular variable neutral density filter (NDF), half wave plate (HWP), high magnification zoom lens (ZL), CCD camera, convex lens (L1, L2, L3), sample (S), beam block (BB), variable diaphragm (VD), polarizer (P), detector (D). (b) The schematic diagram of AL of the Si detector on the divergent light due to the significant lens spherical aberration. (c) The picture of the converging state of divergent SC at the focal point of the paraxial light for the L3 lens system with NA 0.35. (d) The measured relationship between T and NA for the L3 lens system with NA 0.35 in virtue of AL of the Si detector.

## 2.2 The irradiation fluences of the pump pulse and the probe pulse

In the study we focus on the strong-field ionization properties of fused silica with laser fluence approaching the damage threshold. In order to avoid the large signal noise stemming from material damage, realize high-repetition-rate continuous acquisition at a fixed spot, and thus greatly improve the stability and consistency of data acquisition, all the transient spectra are acquired with the pump pulse fluence controlled below the damage threshold. Furthermore, for probing into the critical ionization state of femtosecond laser-induced breakdown, the pump pulse fluence is adjusted carefully to approach the damage threshold in a series of increasing fluence. Actually, in the pump-probe measurements for a series of fluence, along with the pump pulse fluence increasing gradually towards the damage threshold, the sample surface will ultimately be damaged—the situation can be observed directly by the CCD camera monitoring the irradiated surface or detected from the abnormal steep drop of the signal curve for a specific pulse fluence. Accordingly, the damage-threshold fluence ($F_{th}$) is defined as the laser fluence leading to the initial failure of fused silica surface, which is measured to be about 1.8 J/cm$^2$ for the typical 795-nm, 90-fs pulses. Because in the study special attention is paid to the damage threshold, we use normalized laser fluence $F_n = F/F_{th}$ instead of actual laser fluence F to present the fluence dependence of FWPTS consistently and quantitatively, and thus highlight the critical ionization characteristics.

With regard to the probe beam, we let the probe pulse fluence keeps constant and weak enough in the measurements for a series of pump pulse fluence, and thus ensure that the probe pulse do not generate observable effects on the ionization process induced by the pump pulse. In detail, for the maximum pump fluence approaching $F_{th}$, the maximum fluence ratio of pump pulse to probe pulse is about 500:1, and for the minimum pump fluence leading to initial observable effects, the minimum fluence ratio is about 200:1.

*2.3 The light-collecting module sensitive to the divergence state of probe beam*

As schematically shown in Fig. 1(b), the newly-developed light-collecting module focuses on the divergence state of probe beam. Basically, for focusing on the sample surface, after transmitting the sample the probe beam diverges with a base numerical aperture (NA) value about 0.1. Actually, due to TNOEs aroused by the interaction between strong pump pulses and fused silica, the divergence state of probe beam will be changed as a function of the pump-probe delay time, which serves as a important transient signal linking to the instantaneous effects of laser optical field in our pump-probe measurements. For example, the generation of supercontinuum (SC) may lead to the coloured conical emission with a large divergent angle cone (the spectrum and angular distribution of scattered light satisfy phase matching conditions) [62-64]. Above all, for providing the ability to evaluate of the total energy transmittance of probe pulses, the light-collecting convex lens (L3) is setting with NA 0.35 far larger than 0.1 to fully collect the divergent transmitted probe beam (for achieving the NA value, L3 provided with an effective focal length of about 7 cm is composed of two close convex lens with the focal lengths of 10 cm and 20 cm, respectively (the distances from L3 to the irradiated spot on the fused silica sheet and to the Si photodiode detector are about 8 cm and 60 cm, respectively)). Then, it is worth noting that although NA 0.35 is large enough to collect the diverging probe beam without observable loss of light for the aperture effect of L3 (the loss of light by absorption and reflection of L3 is not taken into account here), there is loss for the light with a large incident angle (large NA) in the photoelectric detection process, which is owing to the aperture effect of the Si photodiode detector: because of the significant spherical aberration of the L3 lens system, the light emitted from a single point (that is the irradiated spot) fails to converge in a single focal point (the focal point of the paraxial light and that of the non-paraxial light don't meet in the same location of the optical axis); thus, when the Si photodiode detector with a detector aperture size of Ø9.5 mm is placed at the focus of the paraxial light, the light with a large incident angle will not enter the effective detection aperture; therefore, when the transmitted probe light is strongly diverging for the significant TNOEs, the light-collecting module will naturally produce an aperture limitation (AL) on the divergent light and lead to the descending region in the FWPTS curve.

In order to visualize the effect of spherical aberration of L3 for the strongly-divergent light, under the similar light-collecting condition of L3 as in the pump-probe measurement, we use the divergent SC outputted from a SC source (YSL Photonics, SC95) and diverged by an objective with NA 0.35 instead of the actual divergent probe beam, to imitate the light field distribution near the detection plane, as shown in Fig. 1(c). Due to the strong spherical aberration, significant astigmatism formed by the light with a large incident angle can be observed clearly on the periphery of the paraxial converging focal spot. As a result, the paraxial light converging to the central area can enter the detection aperture of the Si detector, whereas the peripheral astigmatism coming from the non-paraxial light with a large divergence angle is outside the detection aperture and will be lost in the detection.

To quantitatively evaluate the relationship between T and NA for the divergent probe beam collected by the light-collecting module, we further use objectives of NA = 0.1, 0.2, 0.25, 0.35 and a variable diaphragm to obtain a series of divergent beams with different NA values (here the SC output from the SC source is filtered by a 800-nm narrow-band filter with a 40-nm bandwidth and polarized by a polarizer). Based on the series of divergent beams, the intrinsic relationship of T vs. NA for the light-collecting module is measured in Fig. 1(d). Clearly, when NA is greater than 0.225, T will decrease significantly due to AL caused by the strong spherical aberration. Correspondingly, FWPTS will show an obvious T pit for the probe beam with strong transient divergence of NA > 0.225. As described above, the transient divergence of the probe beam with large NA may originate from the conical emission of SC generation, a phenomenon ascribed to TNOEs. In addition, when ROFI is strong enough, OKE will saturate [60,61] and the refractive index variation ($\Delta n$) may turn to a negative value—in such a situation, a transient negative lens (NL) may form in the irradiation region of pump pulse and generate a divergent probe beam, also known as the plasma defocusing [62-64]. For fused silica, as demonstrated by our following experimental results, the strong divergence should mainly be ascribed to TNOEs, which exhibits a near-linear relationship with light intensity (ROFI should have a high-order nonlinearity relationship with light intensity)). Because the optical Kerr respond of TNOEs is instantaneous [54] for a near-100-fs pulse, the cycle-averaged signal of TNOEs should follow the pump-pulse envelope. Therefore, such AL can serve as a simple method to determine the time zero of pump-probe system accurately. Moreover, it opens a window for us to probe into the instantaneous effects of laser optical field in the strong-field nonlinear polarization and ionization of solids.

*2.4 The different settings of the light-collecting and the polarization state*

For the actual wavelength-degenerate pump-probe measurements, more different setups may be employed for focusing on different physical factors and characterizing the respective mechanisms. In order to amplify the TNOE signal for the divergent probe beam with NA < 0.225, the two convex lenses in the L3 lens system with the NA setting of 0.35 may be replaced by an alone convex lens with the NA setting of 0.12. In virtue of the lower NA of the light-collecting module, stronger AL can be obtained for the transiently-divergent probe beam originated in SC generation or NL formation—here more light (outside the light cone with NA 0.12) cannot be collected, and thus a more obvious pit in the T curve should be observed near the time zero.Use initial cap for first word in title or for proper nouns.

For the sake of brevity, in the result part the two general light-collecting settings of high NA 0.35 and low NA 0.12 are also directly marked as "HNA" and "LNA", respectively. In addition, the settings of much lower NA, such as 0.03 and 0.01, can be achieved via placing the variable diaphragm with a specific opening size between the sample and L3.

Besides sensitive to the light-collecting NA, the features of FWPTS of the probe beam are also sensitive to the polarization states of pump and probe beams, which can be changed by the half-wave plates. Actually, there are four typical polarization combination modes for the pump and probe beams: horizontal vs. horizontal ("H+H"), horizontal vs. vertical ("H+V"), vertical vs. horizontal ("V+H"), vertical vs. vertical ("V+V"). For the orthogonal polarization setting, that is, the cases of "H+V" and "V+H", a polarizer oriented along the probe beam polarization may be used in front of the Si detector to magnify TNOEs—these setting cases are denoted as "H+V+P" and "V+H+P".

Moreover, considering the wavelength spectrum of transmitted probe beam to be changed by TNOEs, we measure the spectra of different delay times via the fiber optic spectrometer in the spectral measurement mode of the system. As a matter of fact, a narrow-band filter with a central wavelength of 800 nm and a FWHM bandwidth of 25 nm placing in front of the Si detector can also magnify TNOEs, similar to the effect of the variable diaphragm.

**3. Results and Discussions**

*3.1 FWPTS in the polarization-degenerate setting with different light-collecting NA values*

First, concerning FWPTS in the simplest case, we carry out the pump-probe experiment in the polarization-degenerate setting of H+H. In Fig. 2(a) towards the collecting NA 0.35 (HNA), the transmissivity (T) of the probe beam are presented as a function of the time delay for a series of increasing pump fluences approaching the damage threshold. With regard to the low $F_n$ cases of 0.32 and 0.54, T first slightly descends and then rises in a time scale near the pulse duration. This is a typical curve feature of bipolar shapes associated with the energy transfer of two-beam coupling (TBC) [65-67] of the pump and probe beams, which is a TNOE for the wavelength-degenerate pump-probe setting (here the down-and-up trend of the bipolar shape hints the pulse with weak positive chirp, which should be due to the long transmission distance (over 6 meters) of the pulse in air from the laser source to the pump-probe device). Then, as $F_n$ increases from 0.69 to 0.94 approaching the damage threshold, two recessed regions in the curves become increasing more pronounced, taking place of the alone TBC signal dominated in low $F_n$. The feature indicates two ultrafast physical processes related to distinct mechanisms playing important roles in the damage-threshold strong-field-solid interaction: the first process shows a short duration near the laser pulse duration, with the pit center located in the middle of the TBC process, and the second process shows a long duration evidently larger than the pulse duration, with the pit center lagging hundreds of femtosecond from the first one.

Furthermore, in Fig. 2(b) towards a small NA 0.12 (LNA), similar FWPTS exhibiting two dynamic processes can be obtained with a stronger first one, which almost blankets the TBC effect. As discussed in the experiment part, the light-collecting setting with a lower NA should enhance AL of the divergent light that roots in TNOEs owing to the transient laser field. Here considering the enhancement of the first process in the small NA case, it is naturally to ascribe the first process to such AL of the divergent light occurring in the strong-field-solid nonlinear interaction. That is, the pit center of the first process should be the time zero of the pump-pulse measurement. In addition, with regard to the high time evolution symmetry of the first process relative to time zero, it should be a nearly-reversible process in strong-field-solid interaction from the energy viewpoint [54]. Resembling the characteristics of virtual population observed in the attosecond detections [51-56], this instantaneous, reversible process should be ascribed to the cycle-averaged VOFI that significantly contributes to TNOEs at the near-damage-threshold fluence.

By contraries, besides the longer duration and the delayed response, the second process appears an asymmetric time evolution characteristic: the decline (excitation) stage evolves faster than the rising (relaxation) stage. Moreover, the emergence of the second process is more approaching the damage threshold than that of the first one, meaning a stronger nonlinear characteristic. Referring to the reported pump-probe results about femtosecond laser-induced ionization of solids [29-49], it is easy to arrive at this conclusion that the second process corresponds to the dynamics (excitation and relaxation) of free electrons generated by femtosecond laser-induced ionization, including the initial ROFI and the subsequent II. From the energy viewpoint, the second process can be considered as an irreversible process [54] leading to energy transfer from the light field to the solid. Accordingly, for the sake of brevity, we denote the first process as the reversible process (RP), and the second process as the irreversible process (IP).

To evaluate the amplitudes and durations of the two processes, we use the normal (Gaussian) distribution and the log-normal distribution to fit the two pits in FWPTS, presenting the symmetric and asymmetric features of the two processes, respectively. As shown in Figs. 2(c) and (d), for two typical T curves corresponding to the $F_n$ cases of 0.94 in Fig. 2(a) and 0.85 in Fig. 2(b), such two functions can provide a good fitting of the original data. In detail, both the fitting normal-distribution functions of RPs in Figs. 2(c) and (d) have a FWHM duration of 150 fs, near the FWHM duration of the autocorrelation signal of pump pulses. The results further indicate that RP is directly determined by the instantaneous optical field of the pump pulse and has a significant reversible characteristic. In contract, the fitting log-normal-distribution functions of IPs in Fig. 2(c) and Fig. 2(d) have the FWHM durations of 625 fs and 860 fs, respectively, with the similar delay times about 310 fs to the time zero. The obvious delay between the action peak of the pump pulse and the flourishing of generated free electrons is a strong signal that II plays an important role for near-100-fs laser-induced damage of fused silica—the clear result can serve as a decisive experimental evidence for the long and controversial issue [16-28]. On the other hand, it should be noted that a single log-normal-distribution function can already provide a exact fitting for the excitation and relaxation processes of free carriers, and in the relaxation process T recovers to its initial value in a short time scale near 1 ps—the fast relaxation means the ultrafast trapping mechanism of free carriers attributed to the formation of self-trapping excitons [21,30,31].

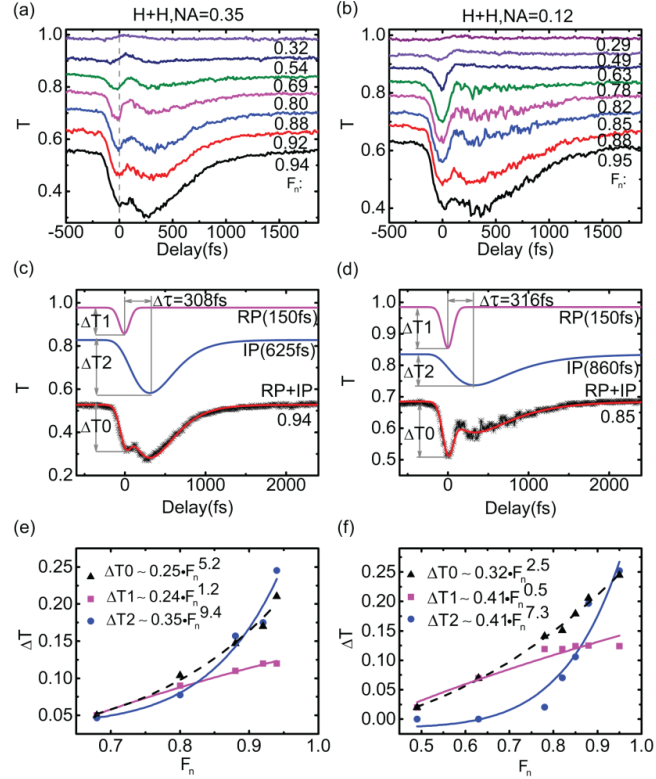

Fig. 2. The transmissivity (T) of the probe beam as a function of the pump-probe time delay for a series of increasing normalized pump fluences ($F_n$) approaching the damage threshold in the H+H polarization setting and the light-collecting settings with (a) NA 0.35 and (b) NA 0.12. In (a), the gray dashed vertical line marks the time zero of the first pit (RP). In (c) and (d), for two typical T curves corresponding to the cases of $F_n$ 0.94 in (a) and $F_n$ 0.85 in (b), the normal-distribution and log-normal-distribution functions are used to fit RP and IP in the respective spectra with the FWHM durations marked in parentheses. $\Delta T0$, $\Delta T1$, and $\Delta T2$ denote the first pit (RP) amplitude measured from the original curve, the first pit (RP) amplitude fitted by the normal-distribution function, and the second pit (IP) amplitude fitted by the log-normal-distribution function, respectively. $\Delta\tau$ denotes the delay time between the two fitted peaks. By fitting the curves provided with obvious pit features in (a) and (b), the dependency relationships between the pit amplitude ($\Delta T0$, $\Delta T1$, $\Delta T2$) and $F_n$ for RP and IP are demonstrated in (e) and (f), which are fitted by the power function with an intercept (the fitted power relations are marked for the corresponding data in the graphs).

By fitting the curves provided with obvious pit features in Figs. 2(a) and (b), we are able to investigate the dependency relationships between the pit amplitude ($\Delta T0$, $\Delta T1$, $\Delta T2$) and the normalized pump fluence ($F_n$) for RP and IP in Figs. 2(e) and (f), respectively, for which the data are fitted by the power function with an intercept. Here, for the first pit, $\Delta T0$ is directly measured from the original data, which would be greatly magnified for the superposed influence of IP with $F_n$ increasing towards 1; ΔT1 is obtained from the fitting normal-distribution function, which is able to offer more accurate evaluation of the amplitude of RP; for the second pit, ΔT2 is obtained from the fitting log-normal-distribution function. In Fig. 2(e), ΔT1 and ΔT2 are proportional to the 1.2 and 9.4 power of $F_n$, and in Fig 2(f), the data are 0.5 and 7.3, respectively (for ΔT0, the power exponent is between the two data for the superposition effect of RP and IP). The data indicate: for RP the power exponent is close to 1, being consistent with the intensity dependence for the main contribution of RP from TNOEs related to VOFI; for IP the power exponent is larger than 7, further confirming the main contribution of IP from II seeded by ROFI. Actually, with $F_n$ approaching 1, ΔT1 seems to appear a saturation trend, which may be due to the saturation of OKE for the plasma effect of ROFI at the onset of optically induced damage [54, 60, 61]; whereas, ΔT2 inclines to a higher nonlinearity, heralding the irreversible destruction of fused silica.

From the results of Fig. 2 for both HNA and LNA settings, it can be confirmed that in the wavelength and polarization degenerate pump-probe measurement, FWPTS of fused silica irradiated by near-damage-threshold, near-100-fs laser pulses always contains two distinguishable physical processes—RP and IP. In Appendix, a more quantitative investigation into the effect of NA values varying from 0.35 to 0.01 is provided, which further confirms the universality of the phenomenon. In general, RP can be monotonously enhanced via decreasing the NA value of the light-collecting module for the enhancement of AL to the divergent probe beam originated in the SC generation. Actually, as shown in Appendix, with $F_n$ extremely approaching 1 IP may also be strengthened for the enhancement of AL to the divergent probe beam originated in NL, in particular for the very low NA 0.01. Interestingly, then the main response signal for IP in detection will turn from the energy change of the probe beam for IBA to the divergence state of probe beam for NL.

*3.2 FWPTS in the orthogonal polarization setting*

It is worth noting that, in above measurements the polarization-degenerate setting of H+H should bring the strongest TNOEs between the pump and probe pulses with the irradiated fused silica as the medium. In the following measurements of Fig. 3 with HNA, the orthogonal polarization setting of H+V+P should greatly suppress TNOEs between the pump and probe pulses [65,66], that is, the RP signal. As shown in Fig. 3(a), at low $F_n$ of 0.43 a TBC process still can be observed clearly, resembling the case in Fig. 2. Then as $F_n$ increases from 0.54 to 0.95, two pits corresponding to RP and IP emerge against. Note that at $F_n$ = 0.95 the RP pit almost merged into the IP pit is weaker than that in Fig. 2, meaning weak TNOEs between the pump and probe pulses in the H+V+P setting. In Figs. 4(b) and (c) for two typical $F_n$s of 0.74 and 0.91, the normal-distribution and log-normal-distribution functions are again employed for fitting the two pits of RP and IP. As the RP+IP superimposed curves shown, the two pits in the measured curves can be fitted well, with a FWHM duration of 250 fs for RP in both two $F_n$ cases, and FWHM durations of 585 and 800 fs for IP in the $F_n$ cases of 0.74 and 0.91, respectively. For RP, here the fitted 250-fs duration is larger than the fitted 150-fs duration in Fig. 2, demonstrating a larger incident pulse duration (determine both the pump and probe pulse durations) in the measurement of Fig. 3 (When the state of the compression grating of the femtosecond amplifier is not optimal, the laser pulse will broaden). Moreover, the larger pulse duration leads to a larger delay time about 490 fs between RP and IP, in contrast to the 310 fs in Fig. 2. On the other hand, for IP, as $F_n$ increases approaching 1 the IP pit expands in time—gradually spreads and covers up the RP pit. For example, with $F_n$ increasing from 0.74 to 0.91, the FWHM duration of IP broadens from 585 fs to 800 fs. Such a phenomenon is consistent with the proposed origin of IP, which is mainly contributed by the II mechanism that produces free electrons in an avalanche process spreading over time [1-15]. In Fig. 3(d), the fitted power exponents for ΔT0, ΔT1, and ΔT2 are 2.5, 0.9, and 3.3, respectively: in the H+V+P setting, the power exponent 0.9 for RP is also close to 1, being agreed with the results in the H+H setting; whereas, the power exponent 3.3 for IP is pronounced lower than that in Fig. 2, which should be due to the larger pulse duration that would reduce the nonlinear exponent of II.

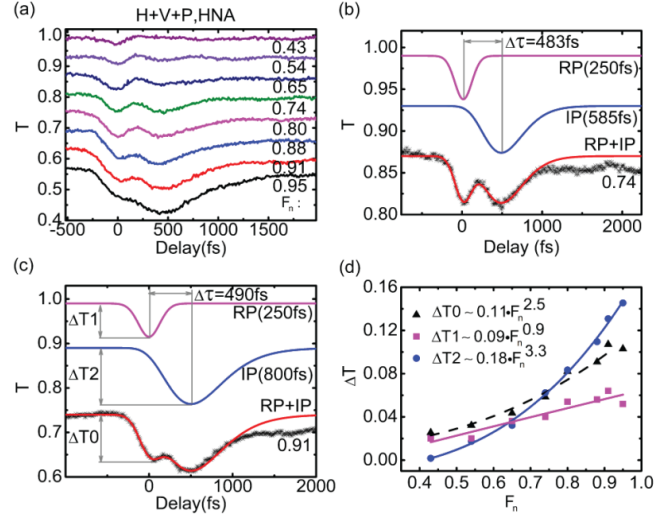

Fig. 3. (a) FWPTS for a series of increasing $F_n$ approaching the damage threshold in the H+V+P polarization and HNA light-collecting setting. For two typical $F_n$s of (b) 0.74 and (c) 0.91, the normal-distribution and log-normal-distribution functions are employed for fitting the two pits of RP and IP with the same definitions of $\Delta T0$, $\Delta T1$, $\Delta T2$, and $\Delta\tau$ as those in Fig. 2. The relationships between the pit amplitude ($\Delta T0$, $\Delta T1$, $\Delta T2$) and $F_n$ for RP and IP are demonstrated in (d), which are fitted by the power function with an intercept (the fitted power relations are marked for the corresponding data in the graphs).

Comparing Fig. 3 with Fig. 2, one can observe that in the H+V+P setting the RP signal is significantly weakened as expected: since the TNOEs between the pump and probe pulses are mostly prevented in such a setting, the VOFI effects induced by the strong pump pulse in fused silica cannot be efficiently detected by the probe pulse. However, note that there is still a weak signal left for RP, which is due to the anisotropic component of the third-order nonlinear susceptibility tensor that has a value of 1/8 of the isotropic one for fused silica towards 100-fs pulses [66]. With respect to the weak signal left for RP, it's going to be an interesting question if we can completely eliminate the RP signal and leave the IP alone. Therefore, to probe into the issue, more systematic comparison measurements aiming at all typical polarization combination settings with HNA are carried out in Fig. 4. In details, Figs. 4(a) and (b) show the spectra measured in the polarization-degenerate (parallel polarization) settings of H+H and V+V, which exhibit the similar two-pit feature resembling that in Fig. 2. Interestingly, in Figs. 4(c) and (d) for the orthogonal polarization settings of H+V and V+H without using the polarizer, the complete dynamic process of IP is present alone in the spectra, without the pit feature of RP. Then, via adding the polarizer, the pit feature of RP returns to the spectra in Figs. 4(e) and (f) for the settings of H+V+P and V+H+P.

These results indicate that, in the orthogonal polarization and HNA settings without using the polarizer, FWPTS is sensitive to the energy reduction of the probe pulse in virtue of IBA, and thus able to detect purely the dynamic process of free electrons excited by the pump pulse via the initial ROFI and the subsequent II. But after a polarizer is employed, the observable energy loss of the probe pulse in RP may occur owing to the reduction of polarization degree of the probe pulse for TNOEs. Actually, such a reduction of polarization degree of probe pulse occurs in both RP and IP, as shown in Figs. 4(g) and (h). However, in RP the reduction appears more obvious for the more significant polarization transformation of TNOEs between the pump and probe pulses with the orthogonal polarization relationship. Like the variable diaphragm used in Fig. 3, the polarizer can play a similar role to enhance RP and IP in the orthogonal polarization and HNA settings. On the other hand, all the spectra in Fig. 4 exhibit a delay time about 250 fs for IP, close to the value of 310 fs shown in Fig. 2.

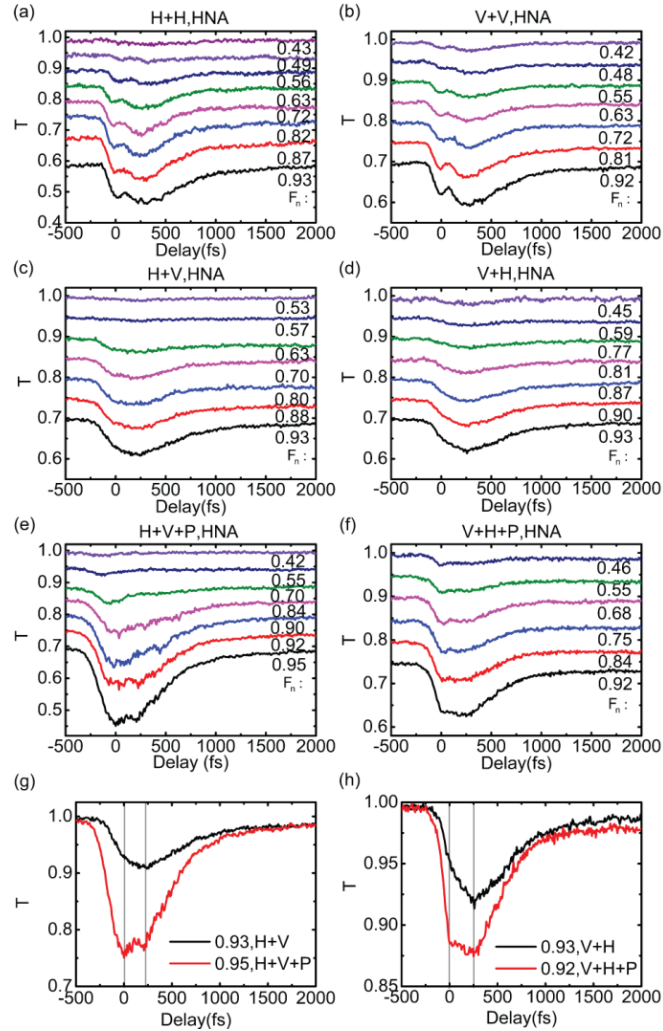

Fig. 4. FWPTS for a series of increasing $F_n$ approaching the damage threshold are presented for all typical polarization combination settings of (a) H+H, (b) V+V, (c) H+V, (d) V+H, (e) H+V+P, and (f) V+H+P with the HNA light-collecting setting. Two typical curves with similar $F_n$ in (c) and (e) (in (d) and (f)) are exhibited together in (g) (in (h)) for comparison (the two gray vertical lines mark the peak positions of RP and IP), which may reveal the effect of the polarizer intuitively.

*3.3 The transient wavelength spectrum of the probe beam*

As above results shown, in the parallel polarization setting RP and IP always appear together, and in the orthogonal polarization setting without a polarizer only IP clearly appears—although in the secondary case the RP signal cannot be detected, it should be clarified that RP still exists as usual: independent of the state of the probe light, the near-damage-threshold strong-field interaction between 100-fs pump pulses and fused silica should always contain the two dominated physical processes of RP and IP, which corresponds to the reversible, instantaneous effects of laser optical field for TNOEs related to VOFI and the irreversible, delayed effects of free electron dynamics related to ROFI and II, respectively. With regard to the clear two-pit feature that can be flexibly controlled in our experiments, such a view is intuitive.

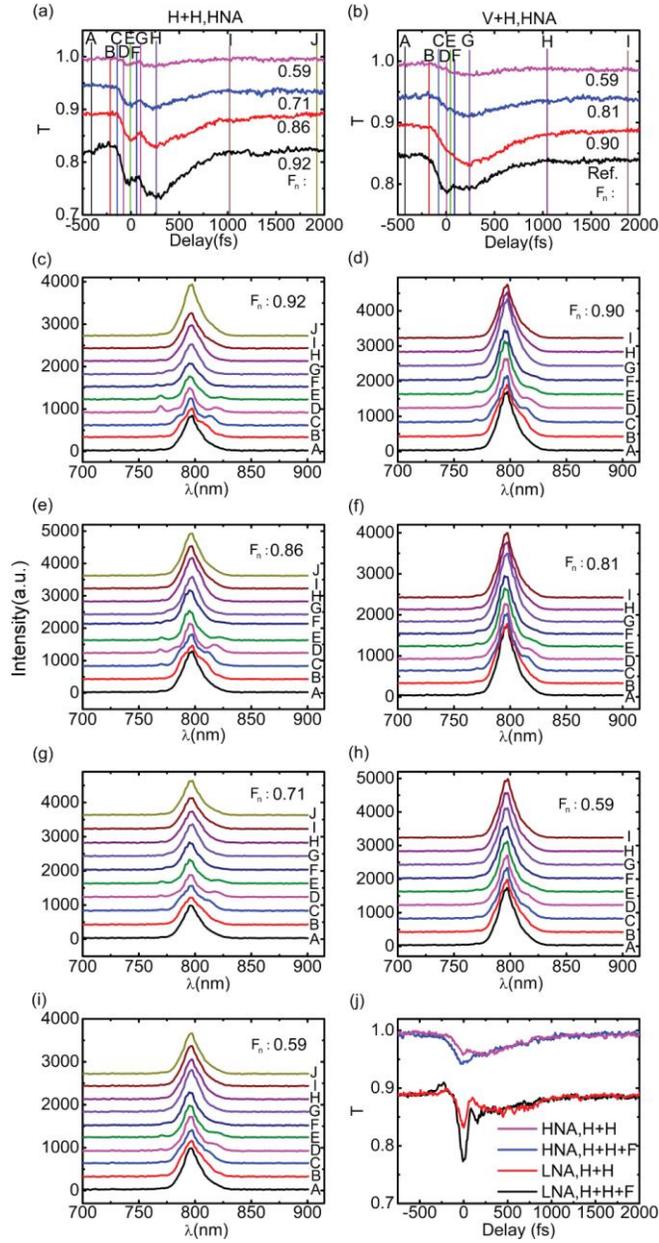

Fig. 5. The transient wavelength spectra of the probe beam for different time delays as marked by capital letters with the corresponding color vertical lines locating the specific times in FWPTS for the experimental settings of (a) H+H, HNA and (b) V+H, HNA. In details, the transient wavelength spectrum series in (c), (e), (g), and (i) are corresponding to the $F_n$ series of FWPTS in (a), and those in (d), (f), and (h) are corresponding to the $F_n$ series of FWPTS in (b) (note that the spectrum marked "Ref." in (b), which is acquired in the H+H, HNA setting, is exhibited here to provide time zero reference for the spectra acquired in the V+H, HNA setting). In (j), the effects of narrow band filters with a central wavelength of 798 nm and a bandwidth of 35 nm placed on the front of the Si detector (marked by "+F" for the corresponding spectrum) are demonstrated for the original settings of HNA, H+H and LNA, H+H.

To further confirm these mechanisms, we measure the transient wavelength spectra of the probe beam for the settings of H+H, HNA and V+H, HNA in Fig. 5. For the H+H, HNA case, the wavelength broadening emerging near the RP pit becomes pronounced with $F_n$ approaching 1. Such a wavelength broadening confirms the main origin of TNOEs for RP, which induce the wavelength conversion and SC generation via self phase modulation, cross phase modulation, stimulated Raman scattering (see the peaks near 770 and 820 nm [68-69]), etc. In addition, free electron plasma generation owing to ROFI occurring in RP may also contribute the wavelength broadening [62]. Note that, for $F_n$s of 0.86 and 0.92, the features of wavelength spectra near the RP pit are already quite similar, which implies a saturation for OKE due to plasma formation. In contrast, for the time range contributed by IP individually (see the position H in Fig. 5(a)), wavelength broadening almost cannot be observed—without the strong-field nonlinear polarization of pump pulses, there is a minimal nonlinear interaction between probe pulses and fused silica. The phenomenon is consistent with the main origin of IBA for the pit formation of IP. In addition, for the V+H, HNA case similar spectral features are revealed with a weaker wavelength broadening effect, due to the far weaker TNOEs felt by the probe pulse for the pump and probe pulses with the orthogonal polarization relationship. The result verifies that even if there is no RP pit in FWPTS, RP still exists, in particular for the strong-field nonlinear interaction between pump pulses and fused silica. Moreover, like the variable diaphragm and the polarizer previously used, narrow band filters are also available to enhance RP, as shown in Fig. 5(j).

## 4. Conclusion

In the paper, FWPTS is developed for probing into the strong-field ionization characteristics of fused silica irradiated by near-100-fs, 795-nm laser pulses with fluence approaching the damage threshold. Besides sensitive to the transient intensity change of probe beam for IBA of free electron plasma generated by ROFI and II, the technique are also highly sensitive to the transient divergence of probe beam for TNOEs (mainly the SC generation caused by OKEs) originated in VOFI from the instantaneous interaction between strong light fields and fused silica. In general, as our results shown, FWPTS exhibits two distinct pits with the short and the long durations, which clearly demonstrates two dominant dynamic processes related to the abovementioned mechanisms in the near-damage-threshold strong-field-solid interaction. In virtue of these spectral features, FWPTS is able to detect the instantaneous effects of laser optical field and the delayed effects of free electron dynamics simultaneously. Thus, it provides a direct and simple way for the comprehensive detections of VOFI, ROFI, and II simultaneously in strong-field nonlinear polarization and ionization of fused silica. In details, the durations, the intensity-dependence amplitudes, and the delay time of the two pits can provide us with the relevant information about the time evolution dynamics and the nonlinear intensity dependence of the two dynamic processes, which serves as the experimental basis for the discussion of proposed theoretical mechanisms. Furthermore, via altering the light-collecting setting and the pump-probe polarization setting, we can flexibly control the amplitudes of these signals, and thus selectively enhance or suppress certain signals to realize the characterization of a specific mechanism. For example, via decreasing the NA value of the light-collecting module, the signal of TNOEs is enhanced obviously, and thus TNOEs highlighted from the background effects of free electron plasma can be investigated explicitly; whereas, in the orthogonal polarization setting without a polarizer, the signal of TNOEs is suppressed greatly, and thus the effects of free electron plasma left alone in the spectrum can be explored separately. Furthermore, the transient wavelength spectra of different delay times reveal the physical relationship between the nonlinear wavelength broadening (shift) and the two dominant processes, which provides more solid experimental evidences for the proposed mechanisms. In short, FWPTS is able to detect the transient dynamics of strong-field nonlinear polarization and ionization of dielectrics with high time accuracy and signal sensitivity, which opens up flexible ways for distinguishing the different mechanisms in the processes.

Actually, from the point of view of energy, the first pit in FWPTS following the laser pulse envelope with a high time symmetry can be defined as RP, indicating the energy-exchange recoverability of nonlinear optical polarization in ultrafast strong-field-solid interaction. Corresponding to the virtual population observed in the current attosecond detections [51-56], this RP related to the instantaneous effects of laser optical field should be ascribed to the cycle-averaged VOFI that significantly contributes to TNOEs at the near-damage-threshold fluence. In contrast, the second pit in FWPTS exhibiting a far longer dynamic process with a distinct delay time of hundreds of femtosecond can be defined as IP, indicating the actual energy exchange from the optical field to the electronic system of the dielectric. Corresponding to IBA observed in previously reported detections, this IP related to the excitation and relaxation dynamics of laser-generated free electrons should be ascribed to the actual ionization including the initial, instantaneous ROFI and the dominant, delayed II that is responsible for the ultimate optical breakdown of fused silica. In general, our study introduces a straightforward method for the simultaneous detection of the two typical transient processes of energy recoverability and dissipation in strong-field nonlinear polarization and ionization of fused silica, which provides deep insights into the dominant ionization mechanism for near-100-fs laser-induced damage of fused silica. For the near-100-fs pulses, our results confirm that II provided with a typical delay time about 300 fs is responsible for the optical breakdown of fused silica. Actually, the femtosecond FWPTS developed in the study has the general significance on a variety of dielectrics not limited to fused silica. Furthermore, in the future systematic FWPTS studies aiming at dielectrics, semiconductors, and conductors would give us a deeper understanding of the universal law of strong-field ionization of solids.

**Appendix**

*The effect of NA values of the light-collecting module on FWPTS*

In the main text, the results in Figs. 2(a) and (b) for two different NA values of the light-collecting module demonstrate that the amplitude of the RP signal may be changed by varying the light-collecting NA. For further quantitatively investigating the effect of NA values on FWPTS, we carry out the transient measurements in the HNA (NA 0.35) light-collecting setting with the NA value varying by a variable diaphragm, as shown in Fig. A1 of the Appendix. Generally, for all the three NA values of 0.35, 0.03, and 0.01, the transient spectra clearly show the RP and IP signals, verifying the universality of the appearance of the two processes for FWPTS. In details, comparing the spectra in Fig. A1(a) and Fig. 2(a) for the resembling light-collecting settings of NA 0.35, one can see that there are similar curve varying trends for the spectra, meaning the similar dynamical processes. It should be noted that in Fig. A1(a) the amplitudes of the signals are greatly reduced with regard to those in Fig. 2(a). This difference should be due to the change of the focusing alignment state between the pump and probe beams (the situation of Fig. A1(a) is not in good focusing alignment state)—actually, it is hard to keep the focusing alignment state completely unchanged in two different measurements after some adjustment of the equipment, and the deviation from the good focusing alignment state may seriously and nonlinearly influence on the signal amplitude nonlinearly. As a result, in Fig. A1(d) the power exponents for $\Delta T0$ and $\Delta T2$ corresponding to RP and IP in Fig. A1(a) also decrease in proportion compared with those in Fig. 2(e).

Then, in Fig. A1(b) along with the NA value decreasing from 0.35 to 0.03, it is obvious that the amplitude of the RP signal increases greatly as expected, resulting in the power exponent for $\Delta T0$ in Fig. A1(e) returning to nearby 1. Whereas, the amplitude of the IP signal decreases in the moderate $F_n$ regime, leading to a rising of the power exponents for $\Delta T2$. As a matter of fact, with the decreasing of the NA value of the probe beam, the sensitivity to the ionization region (that is, the energy change) for the probe beam will decrease correspondingly in virtue of the enlargement of the detection region (the ionization region remains unchanged). This should be an important cause for the decrease of the IP signal in the moderate Fn regime. On the other hand, as described in the experiment part of the main text for AL, a low NA setting is sensitive to the divergence state of the probe beam. Actually, such AL for the low NA setting takes effect not only for RP, but also for IP: when the ionization degree of the irradiated dielectric is high enough, the significant decline of the refractive index will lead to the obvious divergence of the probe beam for NL, and thus the transient spectrum will present a pronounce pit for IP. Such a mechanism resembling the origin for the enhancement of the RP signal, should be the key reason for the rapid increase of IP signal at Fn 0.95. These mechanisms are further confirmed by the transient spectra shown in Fig. A1(c), for which the NA value decreases from 0.03 to 0.01. Here, the signals for RP and IP are both enhanced greatly for AL acting on the divergent probe beam. Due to the notable increase of the amplitude of the IP signal, the power exponent for ΔT0 in Fig. A1(f) is affected by the IP signal again, somewhat like the data shown in Fig. 2(f).

In short, the results in Fig. A1 for a wide range of NA values of the light-collecting module further confirms that in the wavelength and polarization degenerate setting, FWPTS of fused silica irradiated by near-damage-threshold, near-100-fs laser pulses always contains two distinguishable physical processes—RP and IP. With the decreasing of the NA value of the light-collecting module, the amplitude of RP is monotonously increasing for the enhancement of AL to the divergent probe beam originated in the SC generation. Whereas, the effect of NA values on IP appears a bit more complicated. Generally, with Fn extremely approaching 1 IP may also be strengthened for the enhancement of AL to the divergent probe beam originated in NL, in particular for the case with a very small NA value, such as 0.01. Actually, in such a case for IP, the main response signal in detection will turn from the energy change of the probe beam for IBA to the divergence state of probe beam for NL.

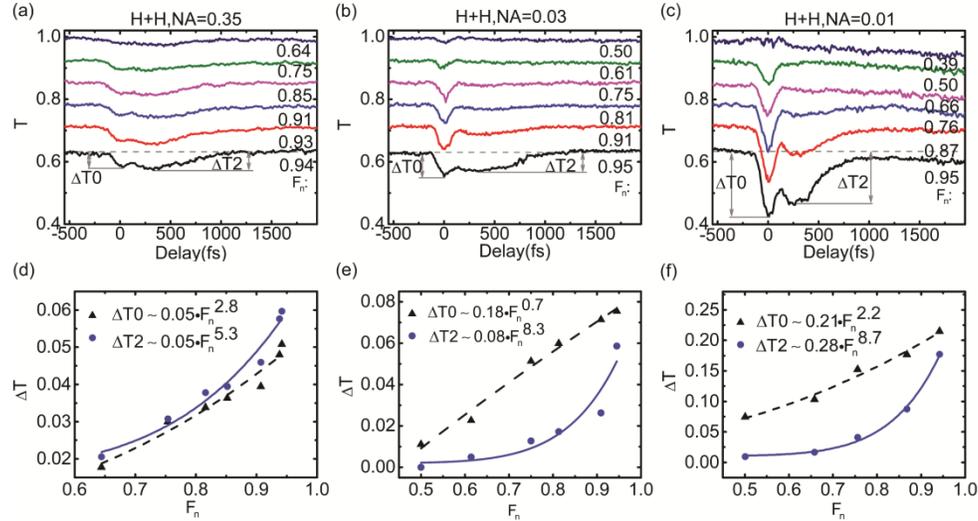

Fig. A1. FWPTS for a series of increasing $F_n$ approaching the damage threshold in the H+H polarization setting and the light-collecting setting with a series of varying NA values of (a) 0.35, (b) 0.03, and (c) 0.01 realized by a variable diaphragm. ΔT0 and ΔT2 denote the first (RP) and the second (IP) pit amplitudes measured from the original curve, respectively. The relationships between the measured pit amplitude (ΔT0, ΔT2) and the normalized pump fluence ($F_n$) for RP and IP are demonstrated in (d), (e), and (f) for the NA value cases of 0.35, 0.03, and 0.01, respectively, which are fitted by the power function with an intercept (the fitted power relations are marked for the corresponding data in the graphs).

**Funding**

National Natural Science Foundation of China (NSFC) (11274400); Pearl River S&T Nova Program of Guangzhou (201506010059); Open Fund of the State Key Laboratory of High Field Laser Physics (Shanghai Institute of Optics and Fine Mechanics); Open Fund of the State Key Laboratory of Optoelectronic Materials and Technologies (Sun Yat-Sen University), and the Hundred Talents Program of Sun Yat-Sen University.

**Acknowledgments**

The authors are grateful to X.R. Zeng for his support in the experiments.